\documentclass[twocolumn,aps,showpacs,floatfix]{revtex4-1}
\usepackage{amssymb,amsmath}
\usepackage{hyperref,graphicx}
\usepackage{dcolumn}
\usepackage{color}
\newcommand{\bs}[1]{{\boldsymbol{#1}}}
\newcommand{\bk}{\bs{k}}
\newcommand{\bq}{\bs{q}}
\newcommand{\br}{\bs{r}}
\newcommand{\bp}{\bs{p}}

\newcommand{\av}[1]{\overline{#1}}
\newcommand{\rmd}{\mathrm{d}}

\newcommand{\intdone}[1]{\int \frac{\rmd #1}{2\pi}}

\newcommand{\tauloc}{\tau_\text{loc}}

\begin{document}

\title{Coherent forward scattering peak induced by Anderson localization}

\author{T. Karpiuk$^{1,2}$, N. Cherroret$^{3,4}$, K. L. Lee$^1$, 
  B. Gr\'{e}maud$^{1,4,5}$, C. A. M\"{u}ller$^1$ and C. Miniatura$^{1,5,6}$}                          
\affiliation{
\mbox{$^1$ Centre for Quantum Technologies, National University of Singapore, 3 Science Drive 2, Singapore 117543, Singapore} \\
\mbox{$^2$ Wydzia{\l} Fizyki, Uniwersytet w Bia{\l}ymstoku, ul. Lipowa 41, 15-424 Bia{\l}ystok, Poland}\\
\mbox{$^3$ Physikalisches Institut, Albert-Ludwigs-Universit\"{a}t Freiburg, Hermann-Herder-Str. 3, D-79104 Freiburg, Germany} \\
\mbox{$^4$ Laboratoire Kastler Brossel, Ecole Normale Sup\'{e}rieure, CNRS, UPMC; 4 Place Jussieu, 75005 Paris, France} \\
\mbox{$^5$ Department of Physics, National University of Singapore, 2 Science Drive 3, Singapore 117542, Singapore} \\
\mbox{$^6$ Institut Non Lin\'{e}aire de Nice, UMR 7335, UNS, CNRS; 1361 route des Lucioles, 06560 Valbonne, France} \\}

\begin{abstract}
Numerical simulations show that,
at the onset of Anderson localization, the momentum distribution of a
coherent wave packet launched inside a random potential exhibits, in
the forward direction, a novel interference peak that complements the
coherent backscattering peak. An explanation of this phenomenon in
terms of maximally crossed diagrams predicts that the signal emerges
around the localization time and grows on the scale of the Heisenberg
time associated with the localization volume. Together, coherent back
and forward scattering provide 
evidence for the occurrence of Anderson localization.  
\end{abstract}


\maketitle

Interference phenomena surviving statistical averages in spatially
random media are of primary importance in order to understand the
transport properties of bulk matter. Already in weakly disordered
media, spectacular deviations from the conventional Boltzmann picture
of transport have been discovered and studied. Important examples
include weak localization and universal conductance fluctuations
in mesoscopic electronic systems, or coherent backscattering (CBS)
and intensity correlations in speckle patterns in the context of wave
transport in disordered media \cite{Montambaux}. In strongly
scattering systems, transport can even be completely suppressed, due
to Anderson localization (AL) \cite{Anderson1958,Houches94}. For
several years now, this phase-coherent inhibition of transport by
disorder has been under active scrutiny in widely different systems
using light \cite{Storzer06}, ultrasound \cite{Hu08}, microwaves
\cite{Chabanov00}, and more recently ultracold atoms
\cite{ColdDisorderRevs,Chabe, Billy08, Roati08, Kondov11,
  Jendrzejewski12}. The  
 observation of AL, however, remains a subtle issue, which requires accurate experiments, with a rather demanding control over absorption and decoherence. 

\begin{figure}
\includegraphics[width=0.9\linewidth]{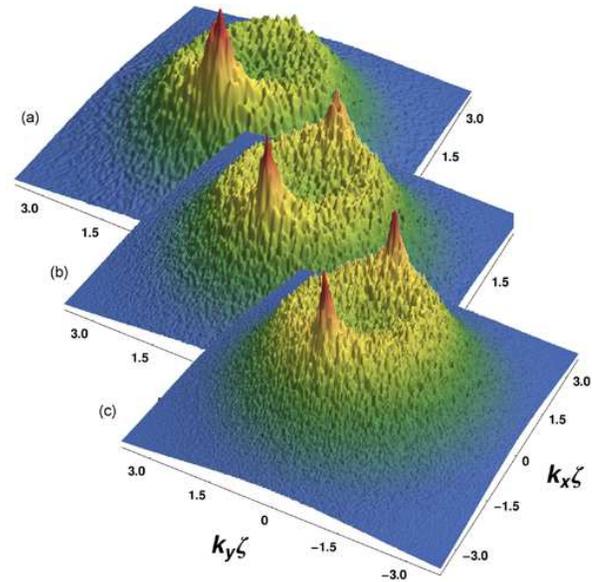}
\caption{(color online) Ensemble-averaged momentum distribution $\overline{n}(k_x,k_y,t)$ of a simulated matter wave launched with initial momentum $\bk_0=(1.5/\zeta,0)$ inside a 2D random potential with correlation length $\zeta$. The time unit is $\tau_{\zeta}= m\zeta^{2}/\hbar$. (a) $t=10\tau_\zeta$: multiple scattering generates a coherent backscattering peak at $-\bk_0$. (b) $t=90\tau_{\zeta}$: a coherent forward scattering (CFS) peak has emerged at $+\bk_0$. (c) $t=950\tau_\zeta$: the CBS and CFS twin peaks in the Anderson localization regime.} 
\label{twinpeaks}
\end{figure}

In order to closely monitor the localization dynamics, we have
recently proposed to study the momentum distribution of a coherent
wave packet launched with finite velocity inside a random potential
\cite{Cherroret12}. Phase-coherent multiple scattering is then clearly
evidenced by the CBS peak, originating from the constructive
interference of counter-propagating scattering amplitudes
\cite{Wolf85,Labeyrie99}. At the onset of AL, the amplitudes cease to
propagate, so that the CBS peak should become stationary, an effect
that we suggested to use as an indicator for AL
\cite{Cherroret12}. The issue proves in fact much richer. 
In this Letter, we report  numerical evidence that AL indeed leads to a
stationary CBS signal, but also triggers the appearance of a novel,
coherent forward scattering (CFS) peak in the opposite, \emph{forward}
direction. 
As a net effect, the momentum distribution shows a remarkable twin-peak structure
(see Fig.~\ref{twinpeaks}). %
We propose a theoretical explanation of CFS via a
combination of the maximally crossed diagrams responsible for CBS.  
This theory implies that the CFS peak, while almost indiscernible for
weak disorder in the diffusive regime, starts to rise around the
localization time $\tauloc = \xi^2/D$ ($\xi$ is the localization
length and $D$ the diffusion constant determined from the initially
measured transport mean free path and time). Furthermore, we find that
the CFS contrast grows on the scale of the Heisenberg time $\tau_H
=h\nu \xi^d$ associated with the localization volume, where $h$ is
Planck's constant and $\nu$ the average density of states per
unit volume. The CBS and CFS twin peaks are thus a clear signal for
the occurrence of AL, and studying their dependence on system
parameters and external fields promises further insights into wave
dynamics in strongly disordered materials.

We consider, then, a quasi-monochromatic wave packet 
that is prepared 
at time $t=0$ with mean wave vector
$\bk_0$ and small spread $\Delta k \ll |\bk_0|$ inside 
bulk disorder. 
In current cold-atom experiments \cite{Labeyrie12, Josse12},
such a wave packet is produced with
ultracold atoms released from a trap inside an optical speckle
field \cite{Clement06}. 
The atoms can be suspended against gravity by a magnetic
field gradient, and accelerated to the desired $\bk_0$ by
a suitable magnetic kick. 
After some time of evolution $t$
inside the disorder, all fields are switched off, and the resulting
momentum distribution $n(\bk,t)$ is accessed 
by time-of-flight imaging.

\begin{figure}
{\centering{\includegraphics[scale=0.20]{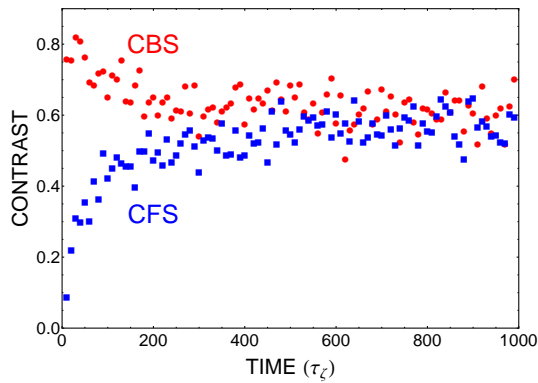}}}
\caption{(Color online) 
Contrast of CBS (red circles) and CFS peak (blue squares) as a function of time, as obtained by fitting the data of Fig.~\ref{twinpeaks} with the theoretical shapes, Eqs.~\eqref{Rho_CBS} and \eqref{Rho_CFS}, and a running average over a time window of 10$\tau_\zeta$.}
\label{figcontrast}
\end{figure}


We first study the matter wave dynamics by numerically solving the
Schr\"odinger equation with Hamiltonian $H=\bp^2/(2m) + V(\br)$ in two
dimensions (2D). 
Without loss of generality, $\av{V(\br)}=0$, where
the overbar denotes the ensemble average over disorder 
realizations. The pair correlation function $\av{V(\br)V(0)} = V^2
C(\br/\zeta)$ defines the potential strength via the variance
$V^2$. It also specifies the spatial correlation length $\zeta$, which
defines a time scale $\tau_{\zeta}= m\zeta^{2}/\hbar$ and energy scale
$E_{\zeta}=\hbar^{2}/(m\zeta^{2})$ \cite{Kuhn07}. 
Fig.~\ref{twinpeaks} shows the momentum distribution
at three different times, for a rather strong potential $V=5
E_{\zeta}$, averaged over 960 realizations of the speckle
potential. The initial conditions are $\bk_0=(1.5/\zeta,0)$ and
$\Delta k=\sqrt{2}k_0/60 \approx 0.0236 k_0$. At short times, in the
diffusive regime [Fig.~\ref{twinpeaks}(a) for $t=10\tau_\zeta$], the
CBS peak at $-\bk_0$ is the only distinctive feature. At longer times
[Fig.~\ref{twinpeaks}(b) for $t=90\tau_\zeta$], a CFS signal has risen
at $+\bk_0$. The CBS and CFS peaks then evolve toward a profile that
remains stationary on the longest times accessible numerically
[Fig.~\ref{twinpeaks}(c) for $t=950\tau_\zeta$]. The peak contrasts,
defined as height over background, are shown in Fig.~\ref{figcontrast}
as a function of time. The rise of the CFS signal goes along with a
reduction of the CBS contrast. In the long run, both contrasts converge
to the same value. Similarly, the widths become comparable, and at the
end of the time evolution covered here, the coherent signals appear as
twin peaks. This progressive symmetrization of the momentum
distribution is consistent with the expectation that it should become
invariant under $\bk\mapsto -\bk$ at very long times because of
time-reversal symmetry \cite{Zimmermann2003}.

\begin{figure}
\includegraphics[height=2.8cm]{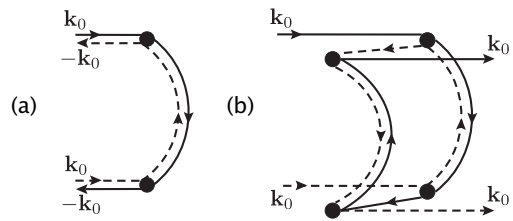}
\caption{Constructive interference of amplitudes along maximally-crossed trajectories, leading to peaks at 
(a) $-\bk_0$ for CBS  and (b) $+\bk_0$ for CFS, respectively. }
\label{cbfs}
\end{figure}

The remarkable symmetry between the two peaks suggests that CFS has the same physical origin as CBS, namely the interference of amplitudes counter-propagating along the same scattering path  
\cite{footnoteBart}. 
This interference contribution is a soft mode of central importance, known as the cooperon in mesoscopic systems \cite{Montambaux}. 
Fig.~\ref{cbfs}(a) represents the real-space trajectories associated with CBS. As shown in Fig.~\ref{cbfs}(b), a CFS signal can appear by concatenating two CBS processes, flipping twice the initial momentum and leading to an enhancement around $+\bk_0$. 
Higher-order CFS (CBS) contributions can arise from chaining an even (odd) number of maximally-crossed diagrams. 
The remaining question then is to understand why the CFS signal, while small in the diffusive regime, is boosted in the Anderson localization regime.   

\begin{figure}
{\centering{\includegraphics[scale=0.50]{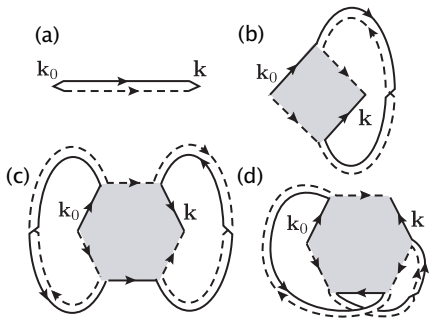}}}
\caption{Ladder (a) and maximally crossed (b) multiple scattering Feynman diagrams, giving rise to the diffusive background and the CBS signal, Eq.\ \eqref{Rho_CBS}. Diagram (c) is responsible for the CFS peak, Eq.\
\eqref{Rho_CFSHika}. Diagram (d) is exactly zero. }
\label{diagrams}
\end{figure}

The quantitative analysis is simplified by assuming a very narrow initial momentum distribution $n(\bk,t=0)=(2\pi)^d\delta(\bk-\bk_0)$ (see below for the impact of a finite momentum spread $\Delta k>0$). The ensemble-averaged momentum distribution at a later time $t$, 
\begin{equation} \label{rhokprimet}
\av{n}(\bk,t) = \intdone{E}\!\intdone{\omega}e^{-i\omega t}\Phi_{\bk\bk_0E}(\omega), 
\end{equation}
is determined by the kernel $\Phi_{\bk\bk_0E}$ of intensity propagation from initial momentum $\bk_0$ to final momentum $\bk$ via energy $E$. This kernel admits a well-known perturbation expansion in the disorder regime \cite{Montambaux}.  
The isotropic diffusive background is described by the series of ladder diagrams represented in Fig.~\ref{diagrams}(a), which results in a maximum background value of $\tau_s/(\hbar\pi\nu) \equiv n_\text{L} $, where $\tau_s$ is the elastic scattering time \cite{Cherroret12}.

The CBS contribution on top of the background is generated by the series of maximally-crossed diagrams of Fig.~\ref{cbfs}(a). Its central ingredient is the diffusive cooperon $P_E(\omega,\bq) = [-i\omega + D(E,\omega) \bq^2]^{-1}$ that peaks around $\bq = \bk_0+\bk=0$. In the diffusive regime $t \ll \tauloc$, $D(E)$ is independent of $\omega$.  The resulting CBS signal then has a decreasing width $(2Dt)^{-1/2}$, because the interfering amplitudes originate from points whose distance grows diffusively \cite{Cherroret12}. In the localization regime $t\gg \tauloc$, this diffusive expansion stops, and the diffusion constant obeys the scaling relation $D(E,\omega) \sim -i\omega\xi^2(E)$ \cite{VW}. Consequently, the CBS signal becomes a stationary Lorentzian  
\begin{equation}  \label{Rho_CBS}
\av{n}^\text{(C)}(\bk) =  \frac{n_\text{L}}{1+\xi^2(\bk+\bk_0)^2}+
\av{n}^{(1)}(k),   
\end{equation}
whose width is determined by the localization length $\xi(E_0)$, evaluated at the effective energy $E_0$ of the excitation with wave vector $\bk_0$ inside the random medium. The second term in Eq.~\eqref{Rho_CBS} is a negative background contribution that ensures particle-number conservation, $\int\rmd\bk\av{n}^\text{(C)}(\bk)=0$. Technically, it is obtained by dressing the cooperon of Fig.~\ref{cbfs}(a) with additional impurity lines \cite{Montambaux}, commonly represented by the shaded 4-point Hikami box of Fig.~\ref{diagrams}(b) \cite{Hikami81}. 

Let us now repeat this analysis for the CFS peak, which originates from the trajectories depicted in Fig.~\ref{cbfs}(b), involving the direct connection of two cooperons. Its contribution to the propagation kernel reads  
\begin{eqnarray}
\label{Phi_CFS}
\Phi_{\bk\bk_0E}^{(\text{CC})}(\omega) && =
\frac{A(\bk,E)A(\bk_0,E)}{(2\pi \nu )^2\,\tau_s}\times\\
 &&\int\frac{\rmd\bq}{(2\pi)^d}A(\bq,E)P_E(\bq+\bk,\omega)P_E(\bq+\bk_0,\omega),\nonumber
\end{eqnarray}
where $A(\bk,E)$ is the spectral function \cite{Kuhn07} and $d$ the spatial dimension. In the diffusive regime, a straightforward calculation shows that the CFS peak has a contrast in 2D of the order of $1/(k_0\ell)$, where $\ell$ denotes the transport mean free path. Thus, CFS is too weak to be visible in the weak-disorder regime $k_0\ell \gg 1$. In the localization
regime, however,  the scaling $D(\omega) \sim-i\omega\xi^2$ makes both cooperons singular and results in a signal that grows linearly in time:  
\begin{equation}  \label{Rho_CFS}
\av{n}^\text{(CC)}(\bk,t) =  3 n_\text{L}
\frac{t}{\tau_H}F_d(|\bk-\bk_0|\xi),  
\end{equation} 
with $F_1(x) =1/(4+x^2)$, 
$F_2(x)=\text{asinh}(x/2)/(\pi x\sqrt{4+x^2})$, and $F_3(x)=\text{atan}(x/2)/(4\pi x)$.   
This CFS signal is indeed an interference peak centered at $+\bk_0$ whose width is again given by the localization length, and whose contrast is of the order of $t/\tau_H$. 

Just as the CBS signal, the CFS contribution should also conserve the particle number. This leads us to dress the double cooperon with the 6-point Hikami box shown in Fig.~\ref{diagrams}(c). This diagram also appears in second-order WL corrections to the electronic conductivity \cite{Hikami81}. As a result of summing 16 different contributions, 
Eq. (\ref{Rho_CFS}) is modified to: 
\begin{equation}  \label{Rho_CFSHika}
\av{n}^\text{(CC)}(\bk,t) =  n_\text{L}
\frac{t}{\tau_H}F_d(|\bk-\bk_0|\xi)
+ \av{n}^{(2)}(k,t),
\end{equation} 
where $\av{n}^{(2)}(k,t)$ is a small negative background contribution that guarantees particle-number conservation, $\int\rmd\bk n^\text{(CC)}(\bk,t)=0$. In principle, there could be other isotropic contributions with an order of magnitude comparable to the CFS signal, such as the one depicted in Fig.~\ref{diagrams}(d), a combination of ladder and crossed series. But this contribution is exactly zero after proper dressing by the Hikami box. Similarly, other possible terms of the same order of magnitude vanish. This clearly identifies the class of diagrams \ref{diagrams}(c) as the one responsible for the CFS peak in the Anderson localization regime. 

We briefly evaluate the effect of a finite source coherence: the convolution of the signal \eqref{Rho_CFSHika} by a source distribution with width $\Delta k>0$ reduces the contrast by a term of order $(\Delta k\xi)^2$. From our numerics for $V=5E_\zeta$, we estimate $\xi\sim8\zeta$ from the width of the CBS peak, such that the reduction remains small, $(\Delta k\xi)^2\approx 0.1$. Having a coherent enough source with $\Delta k\xi\ll 1$ is thus mandatory to observe the CFS peak. In the opposite case, the CBS peak will have already decayed by the time AL sets in \cite{Cherroret12}, now without any remarkable signature on the isotropic background.

\begin{figure}
{\centering{\includegraphics[scale=0.21]{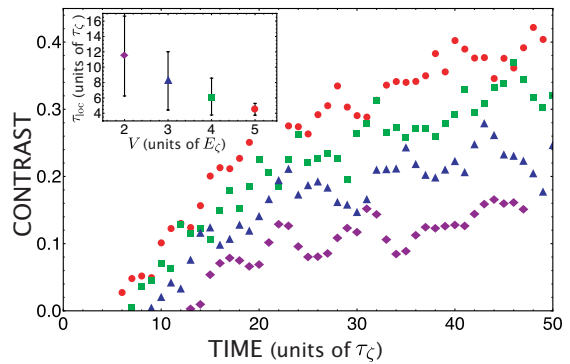}}}

\caption{Early-time CFS contrast for disorder strengths $V= 5 E_\zeta$ (red circles), $V= 4 E_\zeta$ (green squares), $V= 3 E_\zeta$ (blue triangles), and $V= 2 E_\zeta$ (purple diamonds). 
According to our theory, the CFS begins to rise when localization sets in.
The inset shows $\tauloc$, estimated by the earliest CFS appearance, as a function of $V$.}
\label{TDTH}
\end{figure}

Eq.~(\ref{Rho_CFSHika}) implies that from $t\sim \tauloc$ onwards, the CFS peak rises to a contrast of order unity on the Heisenberg time scale $\tau_H= h\nu\xi^d$  associated with the localization volume. Fig.~\ref{TDTH} shows the growth of CFS at early times obtained numerically for different values of $V$. Using the CFS onset as a measure for the localization time, we show in the inset of Fig.~\ref{TDTH} $\tauloc$ as a function of $V$. The observed decrease agrees with the expectation that the localization time is smaller for stronger disorder. For the present parameters, spectral broadening is very important (cf.~the broad background in Fig.~\ref{twinpeaks}). Consequently, the full signal is the sum of contributions from a rather wide range of energies, with lowest-energy components being localized first. A fully quantitative theory for the strong-disorder regime would require to know spectral densities, diffusions constants and localization lengths, which are not yet at hand for spatially-correlated potentials.

Our numerics shows that at long times $t>\tau_H$, the CBS and CFS contrasts saturate at a common value smaller than unity. This dynamics cannot be captured by our theoretical approach because it would require a summation of the full CBS/CFS series of repeated cooperons to all orders, including the dressing by higher-order Hikami boxes, which poses a formidable task. The failure of perturbative expansions beyond the Heisenberg time $\tau_H$ is anyhow well known, and one has to resort to other techniques like supersymmetry or random-matrix theory \cite{Efetov,Beenakker1997}. We expect, however, that the two salient features of our explanation, namely (i) the emergence of CFS around the localization time $\tauloc$ and (ii) a CBS and CFS dynamics on the Heisenberg time scale $\tau_H$, remain valid. Indeed, a similar dynamics was predicted for the dynamical echo within weakly disordered quantum dots \cite{Prigodin94} and observed with ultrasound \cite{Weaver2000}. 
In our setup of bulk disorder, the confinement is due to AL, and the localization length stands for the effective system size. To our knowledge, it is an open question whether the CBS/CFS peaks can remain stationary or perhaps decay to a completely isotropic distribution on very long time scales associated with the glassy dynamics of the localization regime \cite{Mirlin}. 

Future work should also address the impact of finite-size effects on
the CFS signal. In many experiments, waves are sent through a scattering medium with open boundaries \cite{Storzer06, Hu08,Chabanov00}. 
Especially for electromagnetic fields, quasi-plane-wave sources
and detection of radiation patterns in the far field are readily
available, enabling the clear observation of CBS in the reflected intensity 
\cite{Montambaux}. We expect the CFS effect
to be present as well in the transmitted intensity across a strongly
localizing medium, but possibly difficult to observe because this
requires good momentum resolution on the small transmitted signal.

In conclusion, we have discovered a coherent interference phenomenon that is induced by Anderson localization. It manifests itself as an interference peak with high visibility in the forward direction and is closely connected to the CBS peak. 
We have proposed a theoretical explanation by a combination of maximally-crossed diagrams, and predicted CFS to occur in all dimensions $d=1,2,3$. This effect provides an avenue for detailed investigations of AL in strongly random media, and  
should be in reach for current experiments with a good resolution in
momentum space, such as cold atoms,
excitons \cite{Langbein02},
polaritons \cite{Amo09}, and photons \cite{Schwartz2007}.  

We have benefitted from discussions with D.~Delande and V.~Savona, as
well as with T.~Bourdel and V.~Josse and their colleagues.  NC
acknowledges financial support from the Alexander von Humboldt
Foundation and hospitality by the Centre for Quantum Technologies
(CQT), a Research Centre of Excellence funded by the Ministry of
Education and the National Research Foundation of Singapore. ChM and
BG acknowledge funding from the CQT-CNRS LIA FSQL and from the
France-Singapore Merlion program. ChM is a Fellow of the Institute of
Advanced Studies (NTU).

\end{document}